\documentstyle{mn}
\title{On the Mass of M31}

\author[S. T. Gottesman et al.]
       {S. T. Gottesman,$^1$\thanks{E-mail: gott@astro.ufl.edu} J.H. Hunter Jr.$^1$, and V. Boonyasait$^1$\\
        $^1$Department of Astronomy, 211 Bryant Space Science Centre,
            University of Florida Gainesville, FL 32611}

\begin{document}

\maketitle

\label{firstpage}

\begin{abstract}
Recent work by several groups has established the properties of the dwarf 
satellites to M31. We reexamine the reported kinematics of this group 
employing a fresh technique we have developed previously. By calculating the 
distribution of a $\chi$ statistic (which we define in the paper) for the 
M31 system, we conclude that the total mass (disk plus halo) of the primary is
unlikely to be as great as that of our own Milky Way. In fact the $\chi$ 
distribution for M31 indicates that, like NGC 3992, it does not have a massive
halo. In contrast, the analysis of the satellites of NGC 1961 and NGC 5084 
provides strong evidence for massive halos surrounding both spiral galaxies.

\end{abstract}

\begin{keywords}
dark matter -- galaxies: halos -- galaxies: individual (M31) -- galaxies: 
kinematics and dynamics
\end{keywords}

\noindent {\em `There are reasons, increasing in number and quality, to believe
that the masses of ordinary galaxies may have been underestimated by a factor 
of 10 or more' (Ostriker, Peebles and Yahil, 1974).}

\medskip

\noindent {\em `...there is no independent observational dynamical evidence to 
support the view that the masses of galaxies are very much larger than those 
derived in investigations of single galaxies which have been carried out over 
many years' (Burbidge, 1975).}

\section{Introduction}

According to Mateo (1998) more dwarf galaxies have been discovered as Local 
Group members since 1971 than during the preceding 222 years. A large fraction
of these systems are bound to M31. Recently, Evans and Wilkinson (2000, EW), 
Evans et al. (2000) and C\^{o}t\'{e} et al. (2000) have independently estimated
the total mass associated with the M31 (disk plus halo) and found that it is 
probably less massive than our Milky Way. Here we present an alternative way of
estimating the total mass of M31 using these same data but employing a novel 
technique we have developed (Erickson, Gottesman and Hunter 1987, EGH1; and 
1999, EGH2).

The use of the kinematics of dwarf satellite galaxies to investigate the masses
and extents of galaxy halos is a very powerful technique (see EGH1, EGH2, 
Zaritsky et al. 1993, ZSFW1 and 1997, ZSFW2). Almost all studies have been 
limited to describing the average properties of a generic galaxy because very 
few individual systems posses enough satellites to allow unique inferences. 
Owing to the limitations imposed by projection effects, comparisons between 
observations and models (see EGH2, and \S3.1 below) indicate that about seven 
satellites are required before the signature of a massive halo is likely to be 
observed. We will discuss this further below. In EGH2 we studied the statistics
\linebreak of over 71 satellite-primary pairs. Unfortunately, very few of the 
primaries had as many as three satellites. However, NGC 1961 (Gottesman, Hunter
and Shostak, 1983; hereafter GHS) and NGC 5084 (Carignan et al., 1997), as 
individual galaxies, meet the requirements for a substantial number of 
satellites (at least eight for NGC 1961 and seven for NGC 5084). In the Local 
Group, M31 is surrounded by a cloud of dwarf systems (see EW, for example) and 
it is our purpose to examine the kinematics of its satellites using the 
\linebreak methods we have developed in our earlier papers, and to compare the 
properties of M31 with those of NGC 1961 and NGC 5084. In the sections that 
follow, we will briefly review our methodology, discuss the data for the three 
galaxies being compared and draw some conclusions about the properties of M31.

\section{Methods}

We have extended the method introduced by van Moorsel (1982, 1987) for the 
study of binary galaxies to the study of massive primaries with dwarf 
satellites in orbit (greater detail can be found in EGH2). We treat each group
as if it were a set of multiple satellite-primary pairs. We estimate the large
scale mass, $M$, of the spiral (obtained from the orbital properties of each 
satellite, which we will call the orbital mass) in terms of the dimensionless 
function $\chi$ as 

\begin{equation}
 M\chi \equiv r_pV^2_r/G
\end{equation}

\noindent where $r_p$ and $V_r$ are, respectively, the projected radial 
separation, and the difference between the projected radial velocity of each 
satellite and primary. For $r_p$, we use the chord length distance that 
separates the satellite from M31 rather than the arc length. Thus, 
$r_p = 2\,R\,sin(\phi/2)$. $R$ is the distance to the satellite and $\phi$ is
the angle between M31 and each satellite calculated from standard spherical 
trigonometry (Smart, 1977). The data for NGC 1961 and 5084 in this paper were
obtained from a literature search using the NASA/IPAC Extragalactic Database 
(NED) and  differ \linebreak slightly from EGH2.

If each object is idealized as a point mass, bound to the primary (as discussed
in EGH2), the function $\chi$ (which measures the magnitude of the projection 
effects) depends on the true anomaly, $\nu$, the angle between the line of 
nodes and the major axis of the orbit, $\omega$, the inclination of orbital 
plane, $i$, and the eccentricity of the orbit, $e$,

\begin{equation}
\chi = sin^2i\frac{[cos(\nu+\omega)+e\,cos\omega]^2}{1+e\,cos\nu}[1-sin^2(\nu+
\omega)sin^2i]^\frac{1}{2}.
\end{equation}

As an estimate of the importance of any halo we have the observed value, 
$\chi_{obs}$,

\begin{equation}
\chi_{obs} = r_pV^2_r/G(m_1 + m_2)
\end{equation}

\noindent where $m_1$ and $m_2$ are the masses determined from the rotation of 
the individual primary and each satellite respectively. The sum of these masses
is an estimate of the total mass on a small scale. For our systems, $m_2/m_1 
\ll 1$; hence, $m_2$ can be ignored and the subscripts dropped. $\chi_{obs}$ is
related to $\chi$ by

\begin{equation}
\chi_{obs} = (M/m)\chi.
\end{equation}

Thus, if the mass determined from the rotation curve, $m$ (which we will call 
the rotation mass), is the total mass of the galaxy, the two estimates of this
mass will be the same and the distribution of $\chi_{obs}$ will be the same as 
that of $\chi$.

If values for the orbital elements in equation (2) are chosen randomly, the 
theoretical distribution of $\chi$, $P(\chi)$, is found to be sharply peaked 
for small values of $\chi$. That is, there is a strong tendency to 
underestimate the orbital mass, $M$, by almost an order of magnitude, because 
small values of $r_p$ and $V_r$ are highly favored. Furthermore, as long as the
rotation curve reflects all the mass of the galaxy, and the satellites are 
bound, $P(\chi) \rightarrow 0$ as $\chi \rightarrow (1+e)$, which must be less 
than 2. This analysis employs point masses, but in EGH2 we showed that this 
assumption is not critical to the conclusion.

The method has several advantages and we comment on two (see EGH2 for details).
The ratio, $\chi_{obs}$, is independent of the distance scale, and it compares 
directly the total mass of the primary estimated simultaneously at several 
\linebreak different radii. Thus we can separate the notion of a massive galaxy
from that of a galaxy with a massive halo. It is quite certain that the spiral 
galaxies NGC 1961 and NGC 5084 are very much more massive than M31, but that in
itself implies nothing about the existence of heavy halos for any or all of the
systems. We explore this further in the sections that follow.

\section{The Galaxies}
\subsection {NGC 1961 and NGC 5084}

In Figure 1 we show the observed distribution of $\chi_{obs}$, $N(\chi)$, for 
the combined, independent, published sets of data (EGH2 and ZSFW2), augmented 
by five new satellite-primary pairs found by Ellington (2001). In all cases the
projected separations of the satellites are less than 300 kpc (we adopt $H_o = 
67$ km s$^{-1}$ Mpc$^{-1}$). The properties of Ellington's groups are shown
in Table 1. A small fraction, 14 per cent, of the 71 data points have 
$\chi_{obs} > 2$. If contamination by non-physical pairs is not a problem and 
all the satellites are bound, then it is these few points ($\chi_{obs} > 2$) 
that provide information about mass distributions that extend well beyond the 
radius of the disks (halos).

\begin{figure*}
\vspace{97mm}
\includegraphics{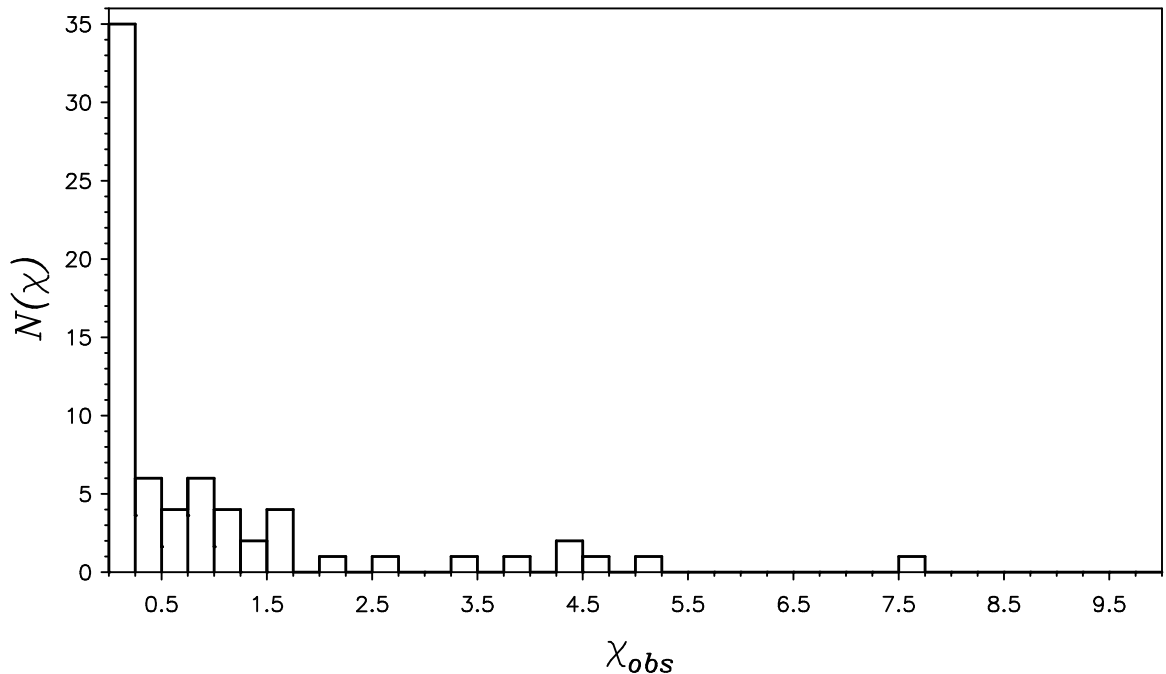}
\caption{The $N(\chi)$ distribution for 71 primary satellite pairs discussed 
         in EGH2 augmented by Ellington (2001). The ordinate is $N(\chi)$; the
         abscissa is $\chi_{obs}$ in units of 0.25. The data are restricted to 
         $r_p \leq 300$ kpc. Ten points show $\chi_{obs} > 2$. One point is 
         off the scale with $\chi_{obs} = 10.6$.}
\end{figure*}

\begin{table}
 \caption{New Primary-Satellite Systems (Ellington, 2001)}
\begin{tabular}{llcccc}\hline
Primary  & Satellites   & $r_p$ &  m  &   M   & $\chi_{obs}$ \\ \hline
NGC 2532 &  CG 0198     &  87   & 16  &  2.9  &     0.18     \\
UGC 1448 & SBS 0155+018 &  43   & 13  &  58   &     4.36     \\
NGC 3677 & CGCG 242-036 &  94   & 92  &  18   &     0.2      \\
NGC 1097 & NGC 1097a    &  16   & 57  &  1.7  &     0.03     \\
         & ESO 416-G032 &  223  & 57  &  3.2  &     0.05     \\ \hline
\end{tabular}

\medskip
{\bf Col. 1:} name of primary galaxy; 
{\bf Col. 2:} name of associated satellite(s); 
{\bf Col. 3:} projected primary-satellite separation, in kpc, 
              based on a Hubble constant $H_o = 67$ km s$^{-1}$ Mpc$^{-1}$; 
{\bf Col. 4:} rotation based mass of primary from HI profile (see EGH2), in 
              units of $10^{10}\;M_{\odot}$; 
{\bf Col. 5:} mass of primary from orbital properties of satellites, in units 
              of $10^{10}\;M_{\odot}$; 
{\bf Col. 6:} the value of $\chi_{obs} = (M/m)\chi$ for each primary-satellite 
              pair. 
\end{table}

\begin{table}
 \caption{Primary Galaxies}
\begin{tabular}{lccccc}\hline
Galaxy   &   D   & $V_{sys}$ & $V(R_{max})$ & $R_{max}$ & Mass ($m$)\\ \hline
NGC 224  & 0.77  &   -301    &     235     &     30    &     38     \\
NGC 1961 &  58   &   3934    &     409     &     36    &    140     \\
NGC 5084 &  30   &   1726    &     328     &     68    &    170     \\ \hline
\end{tabular}

\medskip
{\bf Col. 1:} name of primary galaxy; 
{\bf Col. 2:} with the exception of NGC 224 (M31), distance (Mpc) using 
              heliocentric, systemic velocity (in units of km s$^{-1}$) 
              (Col. 3) corrected to CMB frame and $H_o = 67$ km s$^{-1}$ 
              Mpc$^{-1}$; 
{\bf Col. 4:} rotation velocity measured at maximum radius (in kpc)
              observed (Col. 5); 
{\bf Col. 6:} the resultant mass using a Keplerian calculation, in units of 
              $10^{10}\;M_{\odot}$. 
\end{table}

In EGH2 we made extensive comparisons of our data with those of ZSFW1 and ZSFW2
(ZSFW1 is really a subset of ZSFW2 and we will ignore it in this discussion) 
for which the data were chosen using criteria very different from those used 
in EGH1 and its successors. The question we ask is: what is the probability of 
finding a galaxy-satellite pair with $\chi_{obs}$ greater than two? In EGH2, 
from the analysis of independent data sets, we found that between 13 per cent 
and 21 per cent of the satellites had $\chi_{obs}$ values in excess of two. We 
argued that these differences were not systematic but were measures of 
precision. Furthermore, as we discuss below in \S4.1, the data that 
followed from EGH1 could be modelled marginally by a generic spiral galaxy 
with halo. The details of the model are given in \S4.3 of EGH2. A 
parameter of interest to our discussion here is the eccentricities of the 
satellite orbits which ranged from 0.5 to 0.9. The model, \linebreak  
appropriately defined, was just successful in reproducing the observations, but
qualitatively the initial peak was low and the tail was short. Thus the 21 per 
cent of the values we observed with $\chi_{obs}>2$ is probably at the high 
extreme and the 14 per cent we found for the combined data we show in Figure 1 
is a conservative but not unrealistic estimate. On this basis, we assume a 
probability of one in seven. Thus, primary galaxies must have about seven
satellites before we can have a reasonable expectation of finding one with a 
large value of $\chi_{obs}$. A probability as large as one in five would 
create an even greater problem in consideration of M31, as we will show. (Note,
in EGH2 we discussed a probability distribution, here we discuss counts, 
$N(\chi)$.)

Two galaxies that we have studied that meet this criterion of having at least 
seven satellites are NGC 1961 (GHS, 1983) and NGC 5084 (Carignan et al. 2000).
As we have noted, the essence of our method requires two independent 
methods for estimating the total mass of a galaxy. (1) We must be able to 
measure the rotation velocities of the disk. This provides a mass at relatively
small radii. (2) We must be able to measure the systemic velocities of the 
satellites to the galaxies in question. This enables us to estimate the halo 
masses at larger radii.

For NGC 5084, the properties of the disk were measured by Gottesman and Hawarden
(1986) and the satellites were investigated by Carignan et al. (1997). For NGC
1961, the HI was extensively observed by Shostak et al. (1982). However, the 
nature of the major axis data made it difficult to estimate $V(R_{max})$. EGH2
employed a value consistent with the lower range of the Dutch data. Here, in
addition, we consider the width of the total, or global, HI spectrum. We find
that the maximum velocity lies within a range of 367 and 450 km s$^{-1}$. 
Adopting $V(R_{max}) = 409$ km s$^{-1}$ yields a mass of $1.40 \times 10^{12}\;
M_{\odot}$ (see table 2). This is greater than used in EGH2 and produces 
smaller values of $\chi$.

In Figures 2 and 3 we show the $N(\chi)$ distributions for these two galaxies; 
the
data for all satellites are plotted. In both cases, and more so if the data are
combined, there is clear evidence for the existence of massive halos with sizes
greater than 100 kpc in the case of NGC 5084, and greater than 300 kpc for NGC 
1961. The data for M31 are shown in Figure 4.

\begin{table}
 \caption{NGC 1961 Satellites and Measured $\chi_{obs}$ Values}
\begin{tabular}{lcccc}\hline
Galaxy       & $\Delta$V & $r_p$ &  $M$ & $\chi_{obs}$ \\ \hline
CGCG 329-011 &    +174   &  129  & 91    &     0.65     \\
CGCG 329-009 &     -62   &  152  & 14    &     0.10     \\
LEDA 138826  &    -134   &  188  & 78    &     0.56     \\
UGC 3342     &     +40   &  231  & 8.6   &     0.06     \\
UGC 3344     &    +348   &  342  & 960   &      6.88     \\
UGC 3349 NOTES01) &    +358   &  456  & 1357  &      9.72     \\
CGCG 307-021 &     +67   &  464  & 48    &     0.35     \\
UGC 3349     &    +380   &  522  & 1749  &      12.54     \\
CGCG 329-010 &     +82   &  528  & 82    &     0.59     \\ \hline
\end{tabular}

\medskip
{\bf Col. 1:} name of the satellite galaxy taken from NED; 
{\bf Col. 2:} radial velocity difference (satellite-primary), in units of 
              km s$^{-1}$;
{\bf Col. 3:} projected separation, in kpc;
{\bf Col. 4:} the resultant orbital mass using data in columns 2 and 3, in 
              units of $10^{10}\;M_{\odot}$; 
{\bf Col. 5:} $\chi_{obs} = (M/m)\chi$, the ratio of the mass determined from 
              the kinematics of the satellite in col. 4 to the mass of the 
              primary from Table 2.
\end{table}

\begin{table}
 \caption{NGC 5084 Satellites and Measured $\chi_{obs}$ Values}
\begin{tabular}{lcccc}\hline
Galaxy       & $\Delta$V & $r_p$ &  $M$ & $\chi_{obs}$ \\ \hline
G1  & -40 & 162 & 6.0 & 0.04 \\
G2  & -238 & 97 & 128 & 0.75 \\
G3  & -295 & 71 & 143 & 0.85 \\
G4  & -635 & 94 & 877 & 5.2 \\
G5  & +479 & 51 & 273 & 1.6 \\
G6  & -387 & 61 & 212 & 1.3 \\
$<$G7-G8$>$ & +356 & 134 & 394 & 2.3 \\
G9 & +167 & 103 & 67 & 0.39 \\ \hline
\end{tabular}

\medskip
{\bf Col. 1:} name of the satellite galaxy taken from Carignan et al. (1997);
{\bf Col. 2:} radial velocity difference (satellite-primary), in units of 
              km s$^{-1}$;
{\bf Col. 3:} projected separation, in kpc;
{\bf Col. 4:} the resultant orbital mass using data in columns 2 and 3, in 
              units of $10^{10}\;M_{\odot}$; 
{\bf Col. 5:} $\chi_{obs} = (M/m)\chi$, the ratio of the mass determined from 
              the kinematics of the satellite in col. 4 to the mass of the 
              primary from Table 2.
\end{table}

\begin{figure*}
\vspace{97mm}
\includegraphics{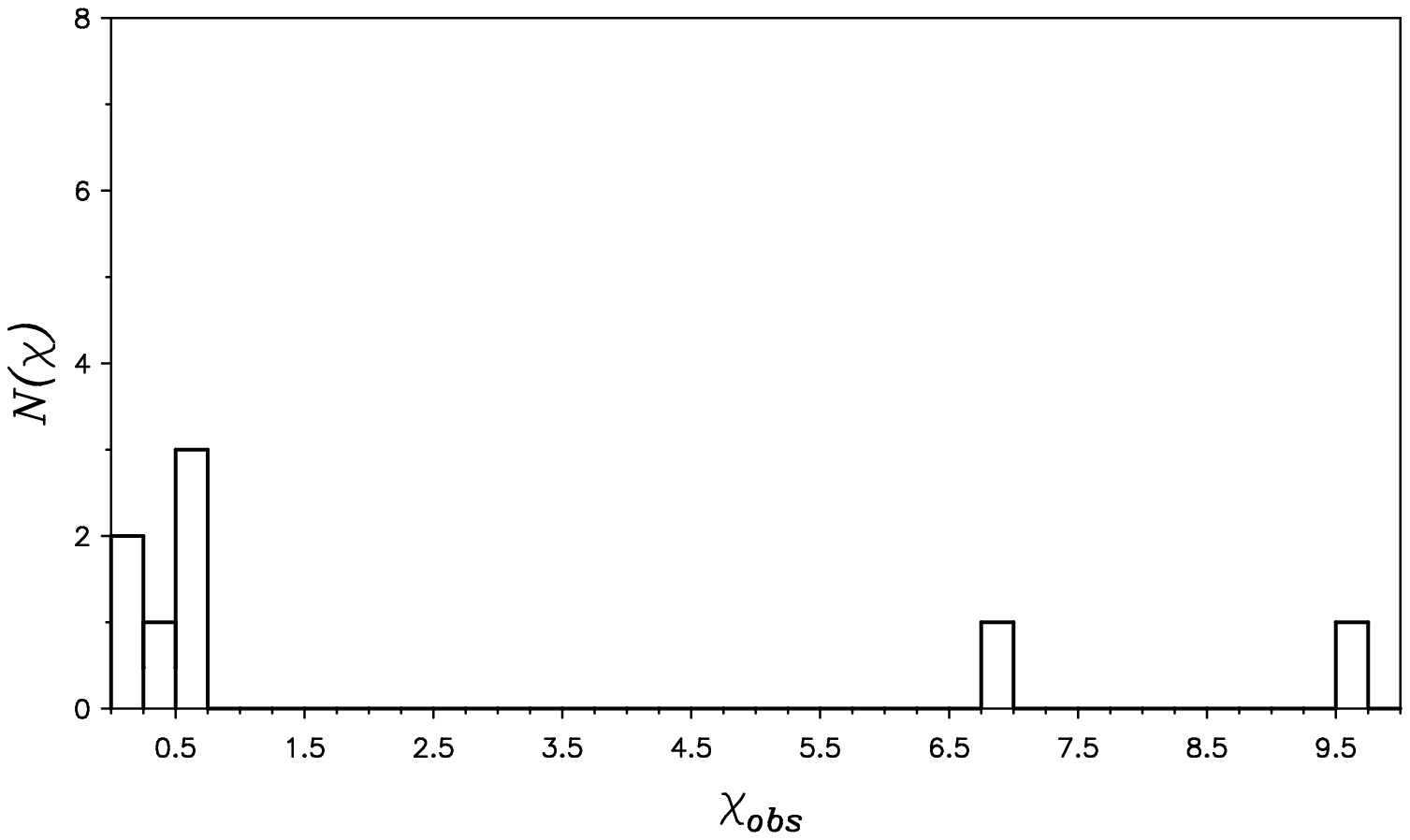}
\caption{The $N(\chi)$ distribution for NGC 1961. The data are taken from 
         Tables 2 and 3. The maximum projected separation for the satellites is
         528 kpc. One point is off the scale with $\chi_{obs}=12.54.$}
\end{figure*}

\begin{figure*}
\vspace{97mm}
\includegraphics{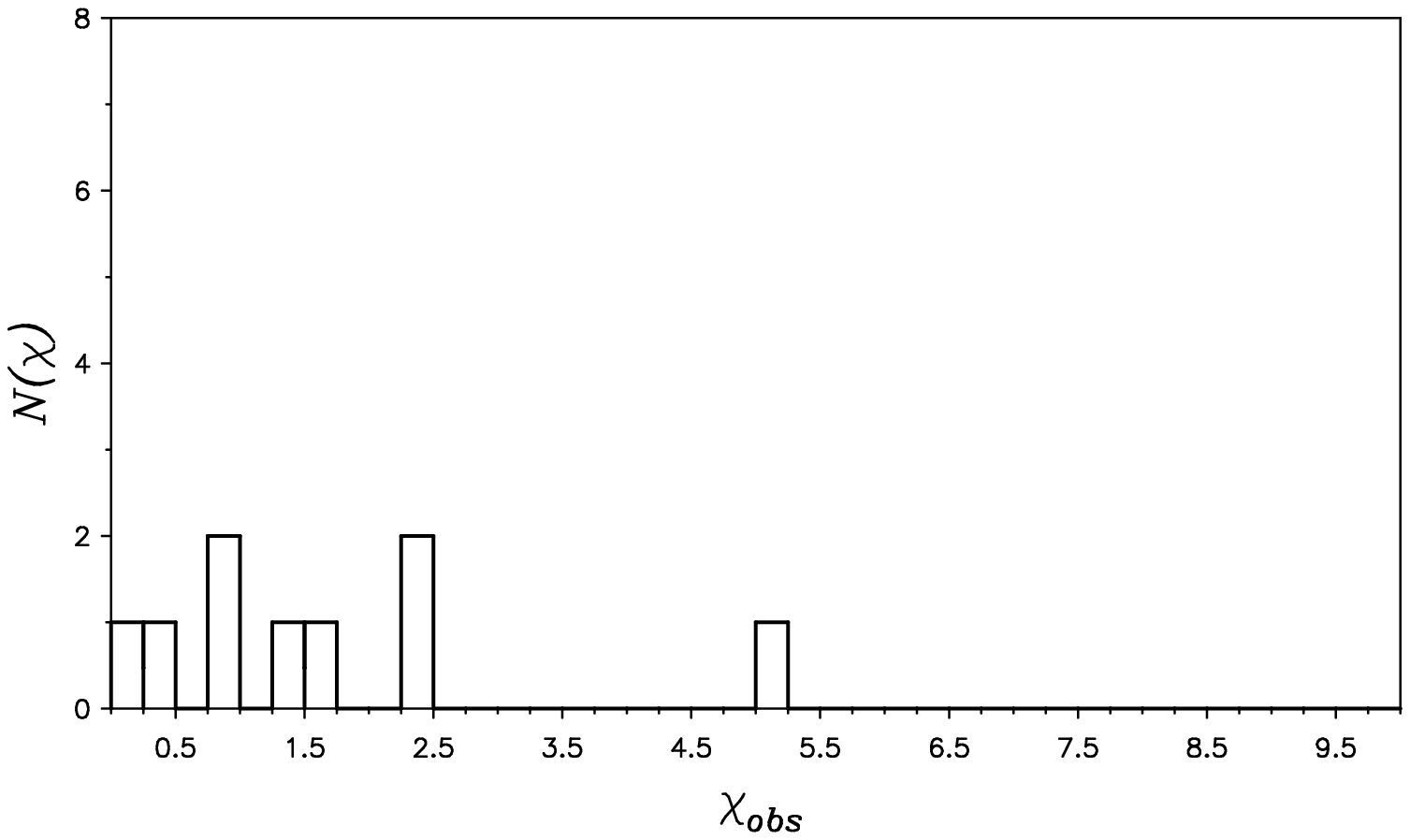}
\caption{The $N(\chi)$ distribution for NGC 5084. The data are taken from 
         Tables 2 and 4. The maximum projected separation for the satellites is
         162 kpc.}
\end{figure*}

\subsection{The M31 Group}

The Local Group is dominated by our own Milky Way and by M31 (NGC 224). They 
are separated by 770 kpc and each is surrounded by a group of dwarf satellites.
Our prime references for the properties of M31 and the M31 group have been
EW, Evans et al. (2000), C\^{o}t\'{e} et al. (2000) and Irwin (private
communication). However, to ensure uniformity our ultimate authority has been
NED, which we use throughout this paper. (Dr. Wilkinson has informed us that
the position given by Evans et al. (2000) for And vii is in error; our values 
are correct.)

In Table 2, we list the properties of the three primary galaxies, while in 
Tables $3-5$ we list the properties of each of the satellite groups. However, 
before considering the $N(\chi)$ distribution for M31 a few comments are 
required.

\begin{table}
 \caption{M31 Satellites and Measured $\chi_{obs}$ Values}
\begin{tabular}{lcccc}\hline
Galaxy       & $\Delta$V & $r_p$ &  $M$ & $\chi_{obs}$ \\ \hline
M32 & +96 & 5.4 & 1.17 & 0.03 \\
M33 & +75 & 198 & 25.5 & 0.67 \\
NGC 147 & +118 & 100 & 32.2 & 0.85 \\
NGC 185 & +107 & 95 & 25.2 & 0.66 \\
NGC 205 & +61 & 8.2 & 0.85 & 0.02 \\ \hline
IC 10 & -33 & 246 & 6.35 & 0.17 \\
$^*$IC 1613 & -60 & 520 & 43.9 & 1.16 \\ 
LGS 3 & -30 & 266 & 5.36 & 0.14 \\
$^*$Peg dwarf & +85 & 412 & 68.4 & 1.8 \\
And i & -85 & 44 & 7.42 & 0.20 \\ \hline
And ii & +82 & 138 & 21.8 & 0.57 \\
And iii & -58 & 67 & 5.3 & 0.12 \\
And v & -108 & 108 & 28.9 & 0.76 \\
And vi & -65 & 264 & 25.7 & 0.68 \\
And vii & +22 & 217 & 2.53 & 0.07 \\ \hline

\end{tabular}

\medskip
{\bf Col. 1:} name of the satellite galaxy taken from NED, IC 1613 and 
              Peg dwarf (Peg) are marked $^*$ as their membership in 
              the group is uncertain;
{\bf Col. 2:} radial velocity difference (satellite-primary), in units of 
              km s$^{-1}$;
{\bf Col. 3:} projected separation, in kpc;
{\bf Col. 4:} the resultant mass using a Keplerian calculation, in units of 
              $10^{10}\;M_{\odot}$; 
{\bf Col. 5:} $\chi_{obs} = (M/m)\chi$, the ratio of the mass determined from the 
kinematics of the satellite in col. 4 to the mass of the primary from Table 2.
\end{table}

\begin{figure*}
\vspace{97mm}
\includegraphics{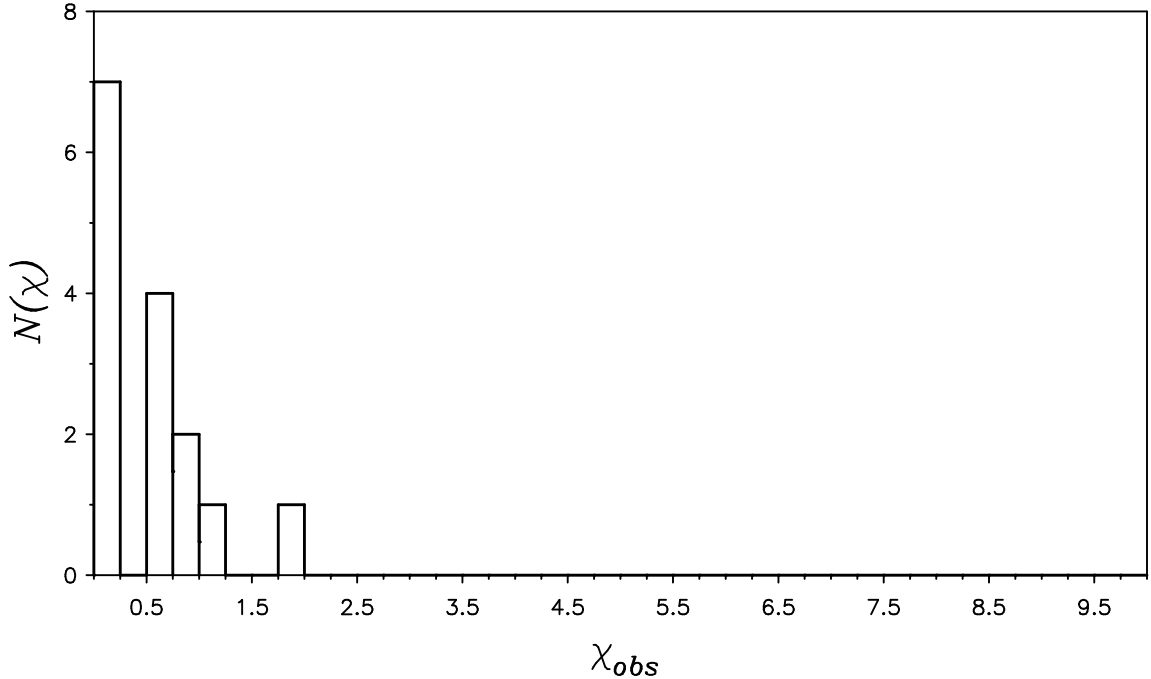}
\caption{The $N(\chi)$ distribution for M31 as listed in Table 5. Data for 
         M31, the primary, is taken from Table 2. The maximum projected 
         separation for the satellites is 520 kpc.}
\end{figure*}

First, what satellites should be included in the analysis? We are concerned in
particular with the Pegasus and IC 1613 dwarfs. These galaxies lie at projected
separations in excess of 400 kpc from M31, which is more than one-half the 
distance to the Milky Way. We presume the orbital period of the satellites to 
be at most a Hubble time. Employing a simple Keplerian calculation, the radius
of the largest orbit is proportional to $M^{1/3}$. If we adopt the mass of NGC 
5084 ($170\times10^{10}\;M_{\odot}$) as a fiducial indicator, the maximum 
radius that satisfies this condition is about 350 kpc. Employing this size may 
be in error, but it will allow us to compare the three galaxies based on the 
initial assumption that all, including M31, are massive systems. It is 
unreasonable to expect a two body calculation to be applicable for satellites 
at such large distances from the primary (M31) in the nearby presence of our 
own massive galaxy. However, as we will show, even if these two distant dwarf 
galaxies are included in the discussion, the M31 group is unusual.

Second, Andromeda is a highly resolved system with an optical diameter in 
excess of three degrees. This makes it unlike any of the other galaxies we have
studied. Our lines of sight to the centre of the galaxy and to the satellites 
are not parallel. Indeed, as independent distances can be measured to the 
satellites, most studies (see for example EW) are done in an Andromedacentric 
system. As noted in  \linebreak equation (3), our formalism has been developed 
for projected velocity and position coordinates. We cannot correct for line of 
sight effects on the observed radial velocity without knowing something about 
space motions. We expect the effects not to be large and certainly to be less 
than factors of three or more required to alter our discussion. We have 
compared a $\chi^{\prime}_{obs}$ calculated using the correct satellite 
distance from M31, as well as $\chi_{obs}$ using projected separations and a 
distance to M31 of 770 kpc. For our data, the $\chi_{obs}$ values from 
$V^2_rr_p$ are about 90 per cent of those obtained from $V^2_rr$.

To gain some insight on this problem it is instructive to consider the special 
case of a circular orbit of radius $r$ and circular speed $V$. In order to 
maximize the $\chi$ values at all orbital phases, we let the orbital plane lie 
perpendicular to the plane of the sky ($i=90^\circ$). The values of $V^2_rr_p$ 
range from 0 to $V^2r$, with the angle averaged values being $4/(3\pi) V^2r =
0.4244\,V^2r$. It should be noted that the coefficient is independent of the 
ratio $r/d$, where $d$ is the separation between the centres of the Milky Way 
and M31. A similar calculation of the angle averaged value $V_r^2r=0.5\;V^2r.$
The coefficients $4/3\pi$ and $1/2$ are maximum values because $\pi/2$ is the 
maximum inclination for any orbit. For this simple example, the angle average 
of $V_r^2r_p$ is roughly 85 per cent of the angle average of $V_r^2r$.

Although it makes no essential difference to our conclusions, we will adhere to
projected separations in order to be as consistent as possible with our earlier
studies. In Figure 5 we show the $N(\chi)$ distribution for the thirteen 
satellites of M31, for which $r_p <$ 350 kpc. This should be compared 
with Figure 6, which shows the combined data for NGC 1961 and NGC 5084 for 
which $r_p <$ 343 kpc. Unlike these latter systems, M31 shows no $\chi_{obs}$ 
values greater than 2. Only if we include the observations for the Pegasus 
dwarf, for which $r_p =$ 412 kpc, do we get a value as large as 1.8 (see Figure 4). Given the 
projected separation, this is surprisingly small, if M31 has a heavy halo. It 
is considerably less than the $\chi_{obs}$ values of five or more seen for NGC 
1961 and NGC 5084. In both of these cases we are considering fewer satellites 
than for M31, and, for NGC 5084 in particular, the satellites, on average, have
smaller projected separations than those observed for M31.

\begin{figure*}
\vspace{97mm}
\includegraphics{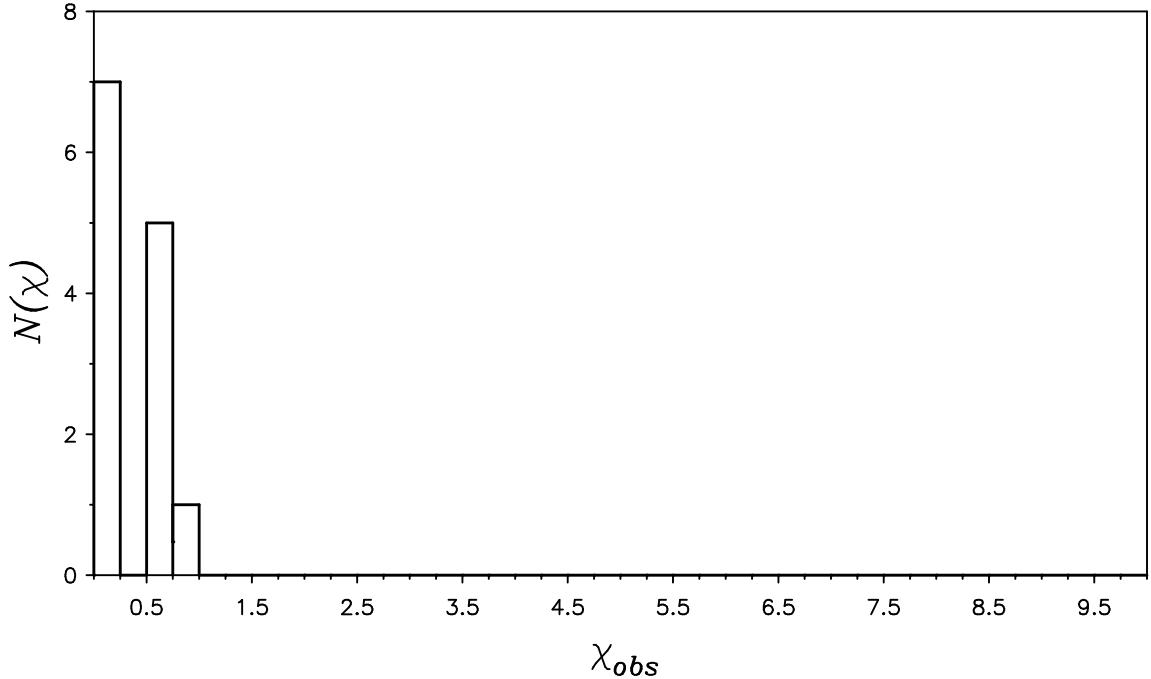}
\caption{The $N(\chi)$ distribution for M31 as in Figure 4 except that the 
         maximum projected separation for the satellites is 266 kpc.}
\end{figure*}

\begin{figure*}
\vspace{97mm}
\includegraphics{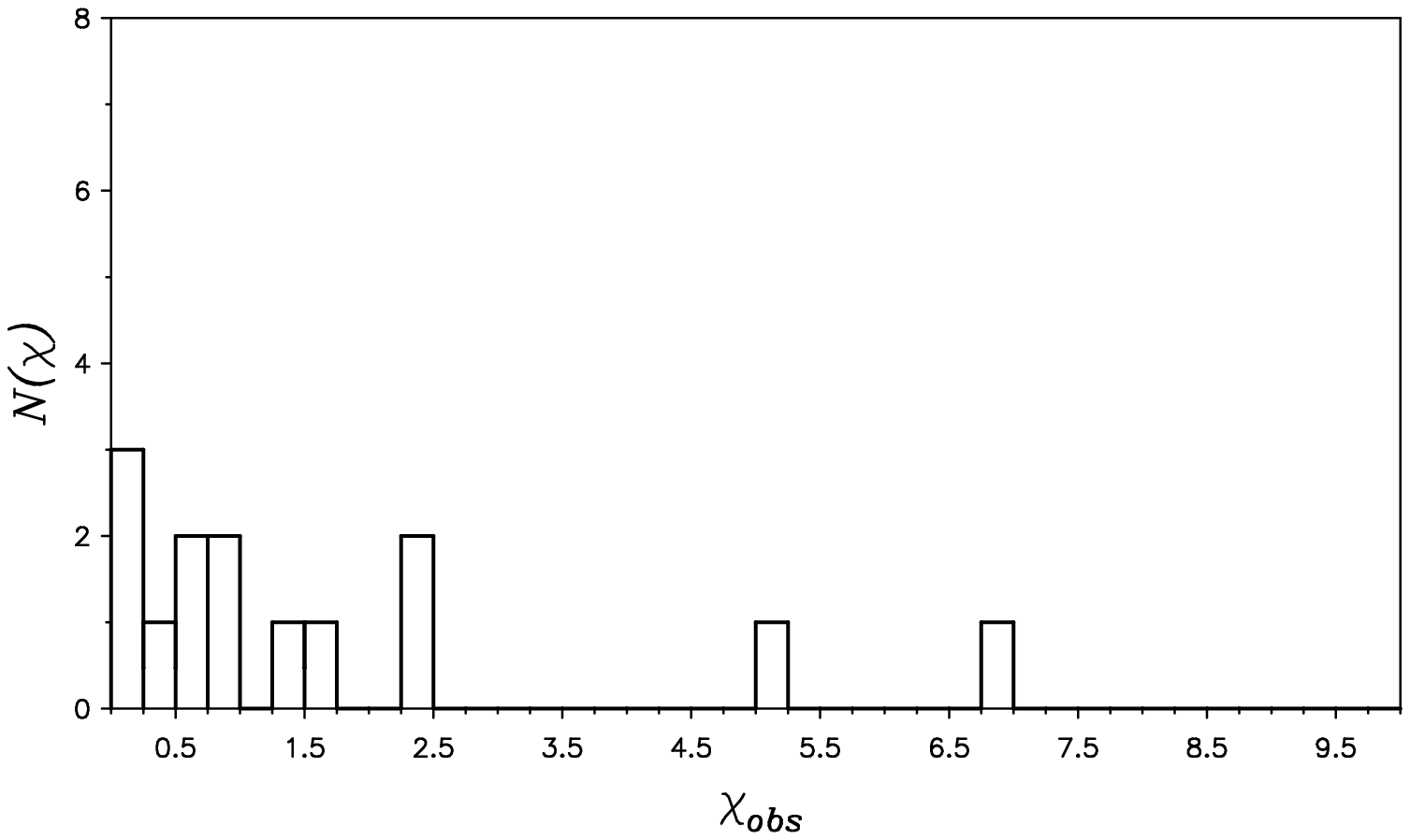}
\caption{The combined $N(\chi)$ distribution for NGC 1961 and NGC 5084. The 
         data are taken from Tables 2, 3 and 4. The maximum projected 
         separation for the satellites is 342 kpc.}
\end{figure*}

The number of galaxies studied by EGH2 was large \linebreak enough that a 
separation could be made between systems having $r_p\leq$ 300 kpc, and those 
having $r_p >$ 300 kpc. The first group of 71 points has been shown in
Figure 1. For the second case, 11 of 37 $\chi_{obs}$ values were
greater than two and of these two had values 10 $<\chi_{obs}<$ 21. Based upon
our studies of bound systems with massive halos, we expect at least 14
per cent (one in seven, as noted in \S3.1) of the satellites to show 
$\chi_{obs}$ values between two and ten, if the projected radius of the group 
is about 300 kpc. None of the satellites of M31 that meet this criterion have 
$\chi_{obs}$ ratios greater than two. Furthermore, if the radius of interest is
extended out to 500 kpc there should be larger values of $\chi_{obs}$, some 
approaching a value of 20. However, $\chi_{obs} = 0.32$ for IC 1613 
($r_p =$ 520 kpc). Thus, the data for M31 differ significantly from the general
galaxies studied in EGH2, and of more interest, the system differs 
substantially from NGC 1961 and NGC 5084, galaxies which we believe show 
conclusive evidence for extended and massive halos. If we consider `satellites'
orbiting at distances as large as 530 kpc, then three out of nine dwarfs 
associated with NGC 1961 have $\chi_{obs}$ values greater than two (6.9, 9.7 
and 12.5). When we include the two points already noted for NGC 5084 (which 
has not been surveyed for satellites with such large orbits), a total of five
satellites out of a population of 17 indicate the presence of a massive 
halo. None of the satellites for M31, out of a population of fifteen, meet this
criterion. The difference is further emphasized by comparing the rms velocity 
dispersions of the satellites to the rotation velocities of the primaries. For 
M31 this ratio is 0.33, while for the combined data of NGC 1961 + NGC 5084 the 
ratio is 0.88 (for all three galaxies, $r_p \leq$ 412 kpc). There appear to be 
two possible conclusions: (1) there is no massive halo associated with M31, or 
(2) the characteristics of the orbits of the satellites to M31, as a group 
(including the effects of dynamical friction), are systematically different 
from those of more isolated `field' galaxies. Either way, the results for the 
M31 system are not consistent with those of other massive galaxies.

\section {Discussion}
\subsection{Kinematics}

In their discussion on determining the mass of spherical systems, Bahcall and 
Tremaine (BT, 1981) employed the kinematical properties of a collection of 
bound particles. They considered an estimator based on the projected mass

\begin{equation}
q\equiv\frac{v^2_rR}{G}
\end{equation}

\noindent where both the velocity and radial separation are projected 
quantities. $q$ serves as an estimator for the mass $M$ of the system. For an 
arbitrary distribution function they showed that the expectation value of $q$ 
becomes

\begin{equation}
<q> = \frac{\pi M}{32}(3-2<e^2>).
\end{equation}

If all the particles have circular orbits $<e^2>\,=\,0$, while for radial 
orbits $<e^2>\,=\,1$. These arguments lead them to define two specialized 
estimators based on the mean of the projected mass $<q>\,=\,<v^2_rR>/G:$

\begin{equation}
M_I = \frac{16}{\pi G}<q>
\end{equation}

\begin{equation}
M_L = \frac{32}{\pi G}<q>.
\end{equation}

For isotropic orbits $<e^2>\,=\,0.5$, and $M_I = M$. For linear orbits $<e^2>\,
=\,1$, and $M_L = M$. Equations $(6)-(8)$ show in a very simple way, but 
clearly, that the estimated mass of M31 depends on the average eccentricity of 
the orbits of the satellites. The greater the typical eccentricity, the greater
the {\it underestimate} of the mass of the galaxy by $<q>$, the estimator of 
the projected mass. From equation (6) the variation between extremes can be as 
much as a factor of three. Of course, the analysis above does not consider the 
influence of a halo with a logarithmic potential.

However, EGH2 did consider the general problem of orbits in massive, isothermal
halos. As the orbits are not closed, the eccentricity was defined, as can be 
done in the Keplerian problem, $e\equiv (r_{apg} - r_{prg})/(r_{apg}+r_{prg})$,
where $r_{apg}$ and $r_{prg}$ are the apogalactic and perigalactic radii 
respectively. In a pure halo potential the large $\chi$ values were found to 
originate in the vicinity of perigalacticon from orbits with large $r_{apg}$ 
(that is, where the apogalactic radius approaches that of the halo, $r_{apg} 
\rightarrow r_H$). In this fashion EGH were able to explain the large $\chi$ 
values of about 20 that were observed in EGH2 as arising from systems with 
massive halos and radii in excess
of 300 kpc. The authors found that, as the eccentricity of the orbits varied 
from 0.90 to 0.00, the maximum value of $\chi$ increased from 9.7 to 24.7. The 
mass of the primary was unchanged for all of these experiments. As the orbital eccentricities increase for a given model, the estimate of the mass (equivalent
to $<q>$) decreases, as was noted by BT. Thus, one way of `hiding' mass, as 
indicated by the $N(\chi)$ distribution, would be to have essentially linear 
orbits. While the galaxy halo would be massive, the orbits would show an 
unexpectedly low value for $\chi_{max}$.

In EGH2 we explored a large number of N-body models designed to simulate a 
range of $N(\chi)$. In order to compare these results with M31 certain 
adjustments to the models are required. The asymptotic maximum velocity has to
be scaled to 235 km s$^{-1}$; the Hubble constant has to be revised to 
$H_o = 67$ km s$^{-1}$ Mpc$^{-1}$, and the characteristic eccentricity of the 
orbits has to be increased at least to 0.8. (See the discussion associated with
\S4 of EGH2.) If the apogalactic distance of the satellites is, for example, 
350 kpc then a larger eccentricity implies that perigalacticon is within the 
disk and the effect of interactions should be severe and clear to the observer.
This does not appear to be the case in general. (However, M32 is a notable 
counter example, which we will discuss below.)

The first model we discuss (model 5 in EGH2) was chosen to reproduce the data 
of EGH1 and its successors. The second model (model 6) was developed to satisfy
the data of EGH2, but did not reproduce the very largest values of 
$\chi_{obs}$. The third model (model 3) was formulated to describe the full 
range of data in EGH2.

{\bf Model 5} was a one-galaxy model, a generic disk plus halo combination. The
halo radius was 90 kpc and the total mass of the system was $9\times10^{11}\;
M_{\odot}$. This model would require many of the satellites of M31 to orbit 
outside the galactic halo.

{\bf Model 6} was constructed to model the data in EGH2. As we will discuss in 
more detail below, the full range of data in EGH2 required a two-component 
model, a very massive disk plus halo constituent, and an essentially haloless 
disk. Model 6 was capable of reproducing the data of EGH2 provided the highest 
values of $\chi_{obs}$ were ignored (possible interlopers). The massive 
component of this model had a halo radius of 225 kpc, and a total mass of 
$2\times 10^{12}\;M_{\odot}$.

In order for the tail of model $N(\chi)$ distributions to approach the largest 
values of $\chi_{obs}$ found in EGH2, model constituents with halo radii in 
excess of 300 kpc were required. {\bf Model 3} was the most successful example.
Two types of galaxy were required to model the full range of $N(\chi)$. One, a 
disk with a massive halo was required to reproduce the high $\chi_{obs}$ tail, 
and two, a disk-only component was necessary to match the peak in $N(\chi)$ at 
small values of $\chi_{obs}$. As in model 6 above, we will assume that the 
massive element duplicates the properties of M31. For this model, the halo 
radius was 350 kpc, and the total mass of the system was $3\times10^{12}\;
M_{\odot}$.

Each of these models was developed independently from numerical experiments 
designed to satisfy different criteria. However, as the data were common and 
several of the characteristics of the models were common (such as the radius of
any disk), it is not surprising that the mass of successful model galaxies 
scales with halo radius. With regard to M31, however, these models have a 
singular failure. The maximum value for $\chi$ is in the range of $9-12$. Such 
models would describe the properties of NGC 1961 and NGC 5084 but not M31. In 
order to approach a maximum value of $\chi_{obs} \simeq 2$, the halo masses of 
the model galaxies must be reduced.

We conclude, it is very unlikely that the mass of M31 within a radius of 350 
kpc is greater than $6\times10^{11}\;M_{\odot}$. This solution requires 
eccentric (near linear) orbits and suggests that the halo dynamics of M31 are 
quite different from those of the galaxies studied by EGH2 and certainly are 
quite different from either NGC 1961 or NGC 5084. 

The data for M31 bear a strong similarity to those of NGC 3992. We have argued
from several perspectives that this galaxy is unlikely to have a massive halo.
(See EGH2 page 158 for further references and for a discussion of the data.)
Like M31, NGC 3992 has a very compressed $N(\chi)$ distribution.

Courteau and van den Bergh (1999) suggest that IC 1613 (the other questionable 
dwarf 
`satellite' of M31, lying at a slightly larger distance from M31 than Peg) is a
free-floating member of the Local Group rather than a satellite of M31. A 
similar argument can be made for the Peg dwarf, given its large separation from
M31 and the nearby presence of the Milky Way. If the radius of any putative 
halo for M31 is closer to 250 kpc, the mass is not greater than $4\times10^{11}
\;M_{\odot}$. Braun (1991), from HI observations, finds a falling rotation 
curve with a value of 200 km s$^{-1}$ at a radius of 26 kpc ($M = 2.4\times
10^{11}\; M_{\odot})$. No heavy halo is required to reproduce his observations.
However, if the rotation curve remained flat out to a distance of 260 kpc (for 
which there is no observational evidence), the implied mass of an isothermal 
halo for M31 would be $2.4\times10^{12}\;M_{\odot}$. Such a system is too 
massive to explain the nature of the $N(\chi)$ distribution for the satellites.
To conclude this part of the discussion, we emphasize that it is very difficult
(unless nature conspires) to have a galaxy system (M31 + satellites) with a 
total mass larger than $4\times10^{11}\;M_{\odot}$, and not have several 
satellites with $\chi_{obs} > 2$ and $\chi_{max}$ substantially greater than 
1.8. If M31 has such a large halo mass, the kinematic properties of the 
satellites should resemble those of NGC 1961 and NGC 5084, which they do not.

\subsection {Dynamical Friction}

EGH2 included dynamical friction in the model calculations, \linebreak and its 
effects cannot be ignored if the satellite mass $m_2 > 10^9\;M_{\odot}$. In 
Model 6 of EGH2, $N(\chi)$ distributions for satellites having masses of 
$m_2 = 2\times 10^9\;M_{\odot}$ were compared to distributions for satellites 
of $m_2 = 0.67 \times 10^9\;M_{\odot}$. After integrating the orbits of the 
higher mass satellites for one Hubble time, dynamical friction was shown to 
truncate the tail at large $\chi$ values, and to increase the peak of the 
distribution at low values of $\chi$ . In other words, for the galaxies with 
massive halos studied by EGH2, the satellites cannot have masses much larger 
than $10^9\;M_{\odot}$ or they will spiral into the nucleus of the primary. One
uncertain parameter controlling deceleration caused by dynamical friction is 
the Coulomb logarithmic coefficient $\Lambda$, which was taken to be 6.8. One 
can define an effective satellite mass $M_{sat}\Lambda$, which in our studies 
was $10^9\;M_{\odot}$, and this value was utilized as the mass of the 
satellites in our numerical experiments. If M31 has a massive halo the 
satellites must have small masses, as defined above. If they have large masses 
the existence of a massive halo would be very doubtful. Unfortunately, this is 
not a sensitive test as most of the satellite masses (that have been measured) 
are significantly less than $10^9\;M_{\odot}$ (Mateo, 1998), with the notable 
exceptions of IC 10, M32 and M33. The implications for the two groups of 
satellites need to be considered separately.

For the majority of satellites, the masses are too small to be effected by 
dynamical friction and cannot be important in truncating the high $\chi_{obs}$ 
tail of $N(\chi)$. If M31 has a substantial halo, the nature of the orbits, and
not dynamical friction, is responsible for diminishing the tail in the 
distribution.

For the massive satellites, the implications are different and interesting. M32
is a special case. It is very close to M31 (the radial separation is estimated 
to be 5 kpc) and interactions are important (for example Byrd, 1976; Bekki et 
al., 2001). At this small separation, this satellite may be about to disappear
owing to tidal and frictional effects.

In contrast, both IC 10 and M33 lie at distances from M31 greater than 190 kpc.
At these distances the frictional effect of the halo is unlikely to be large. 
If the orbits of these systems are deeply plunging (as we have strongly argued
may be the case, if the halo mass of M31 is large) then dynamical friction is
important and the orbits should have decayed over a Hubble time, but they have
not. M31 cannot have a very massive halo, unless the orbits of these massive 
satellites have a low eccentricity. However, if the orbits of all of the 
satellites are near-circular, then $N(\chi)$ is not consistent with the 
existence of a massive halo. Of course, the eccentricity of the satellite 
orbits could change systematically with apogalactic distance, with more 
circular orbits at larger radii. Such complications are not required to 
explain the properties of satellite orbits in the general case, as shown by 
EGH2. Thus, if $<e^2>\, \approx 0$, the M31 system would be most unusual.

It is clear that the majority of the satellites are not massive enough to have 
suffered significant frictional deceleration regardless of the properties of 
the halo. The orbits of the very few (known) massive satellites should have 
suffered from dynamical friction. If they have eccentric orbits they should 
have decayed and now resemble that of the Large Magellanic Cloud (see EGH2) and
be on their last few circuits before oblivion. On the contrary, they lie at 
large radial distances from M31. Either dynamical friction is insignificant or
their orbits are not deeply plunging.

We conclude that: (1) the masses of most of the satellites are too small to 
have suffered a significant deceleration in a massive halo; (2) the massive and
distant satellites may not be on highly eccentric orbits if a massive halo 
exists; or (3) most simply, a massive halo does not exist and frictional 
decelerations are not expected for any of the satellites without regard to 
location or mass.

Mass and length are directly proportional to the value of $H_o$ employed in the
calculations. The results, taken from our models and quoted here, have been 
scaled to $H_o = 67$ km s$^{-1}$ Mpc$^{-1}$. Our models were developed in EGH2 
using a Hubble constant of $H_o = 100$ km s$^{-1}$ Mpc$^{-1}$. However, we 
investigated the characteristics of these model galaxies for different values 
of $H_o$, and no significant differences in our conclusions were found for $50 
< H_o < 100$ km s$^{-1}$ Mpc$^{-1}$.

\section {Conclusion}

In conclusion, we emphasize that the models for M31, that we have considered, 
were contrived to produce the greatest halo mass consistent with the results of
EGH2. We find no convincing evidence for the existence of a massive halo, nor 
do we require unusual characteristics for the orbits of the satellites.  
{\it `Essentia non sunt multiplicanda praeter necessitatum'} (see for example 
Ockham, W. as printed in 1495). 

On the basis of the analysis we have presented, we believe that the Keplerian
mass, consistent with Braun's 1991 observations of M31, is close to the total 
mass of the galaxy. Certainly, if NGC 1961 and NGC 5084 are typical of galaxies
with massive halos, then M31 is not of this class. The $N(\chi)$ distribution 
is too compressed. Our limit is consistent with, but at the lower bounds of, 
the models presented by EW and C\^{o}t\'{e} et al. (2000). However, we are 
dealing with a small number of satellites and caution is required before 
drawing definitive conclusions. The detection of two or three new satellites of
M31, with separations less than but velocities relative to M31 greater than 
that of the Peg dwarf, might alter our conclusions. In addition, the velocities
of the satellites must be made more secure. However, in the absence of such 
data we are in complete sympathy with the suggestion that the Milky way is the 
massive member of the Local Group (Wilkinson and Evans, 1999).

\section*{Acknowledgments}

We benefited greatly from exchanges of email with Drs. Eva Grebel, Mike Irwin,
Mark Wilkinson and from conversations with our colleague, Dr. Ata Sarajedini. 
All were very willing to share fundamental information about the satellites of 
M31. We also appreciate the continued interest shown by Dr. Lance Erickson, and
the efforts of Mr. Chad Ellington in helping with the diagrams and for 
providing some of the data incorporated in Figure 1. Dr. Haywood Smith very 
kindly read critically an earlier version of this manuscript. \linebreak In addition, 
the comments of the referee were very useful in clarifying several points. 
Finally, it would have been much more difficult to develop this paper without 
NED, the NASA/IPAC Extragalactic Database. This is operated by the Jet 
Propulsion Laboratory, California Institute of Technology, under contract with 
the National Aeronautics and Space Administration.

\end{document}